\documentclass[a4paper,11pt]{article}
\usepackage{pos}

\usepackage[symbol]{footmisc}
\usepackage[english]{babel}
\usepackage{graphicx}
\usepackage{graphics}
\usepackage{braket}
\usepackage{bbold}
\usepackage{amsmath}
\usepackage{nicefrac}
\usepackage{dcolumn}% Align table columns on decimal point
\usepackage{bm}% bold math
\usepackage{slashed}
\usepackage{datetime}
\usepackage{mciteplus}
\usepackage{multirow}
\usepackage{siunitx}
\usepackage{booktabs}
\usepackage{color, soul}
\usepackage[usenames,dvipsnames]{xcolor}
\usepackage{float}
\usepackage[utf8]{inputenc}
\usepackage[normalem]{ulem}
\usepackage{mathtools}
\usepackage{setspace}
\usepackage{comment}
\renewcommand{\thefootnote}{\fnsymbol{footnote}}

\newcommand{\drv}{{\rm d}}

\newcommand{\LQCD}{\Lambda_{\rm QCD}}

\newcommand{\Jps}{J/\psi}
\newcommand{\Yps}{\Upsilon}
\newcommand{\ecs}{\eta_c}
\newcommand{\ebs}{\eta_b}

\newcommand{{\HFNRevo}}{\tt HF-NRevo}

\title{Heavy-Flavor Fragmentation from HF-NRevo: \\ Status, Prospects, and Intrinsic Charm}
\ShortTitle{Heavy-Flavor Fragmentation from HF-NRevo}

\author*[a]{Francesco Giovanni Celiberto}
\author[a]{Francesca Lonigro}

\affiliation[a]{Universidad de Alcalá (UAH), Departamento de Física y Matemáticas, Campus Universitario, \\ Alcalá de Henares, E-28805, Madrid, Spain}

\emailAdd{francesco.celiberto@uah.es}
\emailAdd{francesca.lonigro@uah.es}

\abstract{
We report on recent developments of the Heavy-Flavor Non-Relativistic evolution ({\HFNRevo}) scheme, a framework designed to describe heavy-hadron formation through leading-power fragmentation at moderate and large transverse momentum. 
The approach combines short-distance inputs obtained from next-to-leading-order NRQCD calculations with collinear scale evolution in a variable-flavor-number scheme, ensuring a consistent treatment of heavy-flavor thresholds and partonic hierarchies. 
Within this setup we have constructed the {\tt NRFF1.0} family of fragmentation functions for $S$-wave heavy quarkonia in their leading NRQCD Fock states.
We discuss prospective applications of the {\HFNRevo} framework in the heavy-ion environment, where it can provide a perturbative baseline for investigating medium-induced modifications of heavy-flavor fragmentation. 
Its explicit treatment of partonic channels and heavy-flavor thresholds makes it particularly suitable for exploring jet-quenching sensitivity, energy-loss mechanisms, and the emergence of medium-modified fragmentation patterns in the quark-gluon plasma.
The {\HFNRevo} scheme has also been extended to the exotic sector through the {\tt TQ4Q1.x} and newly released {\tt TQ4Q2.0} fragmentation sets, which describe the formation of fully heavy tetraquarks in multiple quantum configurations. 
These developments open a novel pathway to study quarkoniumlike states and to probe the intrinsic charm content of the proton in forward hadron-collision environments. 
Altogether, this program broadens the phenomenological reach of heavy-flavor fragmentation studies at the HL-LHC and future collider facilities, opening access to previously unexplored aspects of QCD and potential portals to New Physics.
}

\FullConference{
Proceedings of the Corfu Summer Institute 2025 \\
``School and Workshops on Elementary Particle Physics and Gravity'' \\
27 April - 28 September 2025 \\
Corfu, Greece
}

\begin{document}
\maketitle

%%%----------------------------------------
\section{Introductory remarks}
\label{sec:introduction}
%%%----------------------------------------

In the exploration of fundamental interactions, hadrons containing open or hidden heavy flavors play a central role as precision probes.
Heavy quarks are particularly relevant in searches for New Physics, as many extensions of the Standard Model predict enhanced couplings to heavy fermions.
Their masses lie well above the QCD confinement scale, which allows perturbative methods to be applied to several stages of their production dynamics.
Bound states of heavy quarks, collectively known as quarkonia, are therefore often regarded as the ``hydrogen atoms'' of QCD~\cite{Pineda:2011dg}.
These systems provide a unique interface between high-precision perturbative calculations and the investigation of the internal structure of the proton.
For example, hadronic decays of $S$-wave bottomonia enable accurate determinations of the strong coupling $\alpha_s$~\cite{Brambilla:2007cz,Proceedings:2019pra}, while measurements of forward quarkonium production constrain the positivity of gluon parton distribution functions in the low-$x$ and low-$Q^2$ region~\cite{Candido:2020yat}.
Quarkonium production is also an important tool for the three-dimensional imaging of the proton, both in the small-$x$ domain~\cite{Hentschinski:2020yfm,Celiberto:2018muu,Bolognino:2018rhb,Bolognino:2021niq,Celiberto:2019slj,Silvetti:2022hyc,Bonvini:2026cxp,Kang:2023doo} and at moderate values of~$x$~\cite{Boer:2015pni,Lansberg:2017dzg,Bacchetta:2020vty,Bacchetta:2024fci,Celiberto:2021zww}.
In addition, unresolved photoproduction of $\Jps$ in association with a charm jet at the EIC has been proposed as a sensitive probe of the intrinsic-charm valence component of the proton~\cite{NNPDF:2023tyk,Flore:2020jau}.
Despite this broad phenomenological relevance, the mechanism underlying quarkonium hadronization remains only partially understood.
Several theoretical approaches have been proposed, but none has yet succeeded in providing a comprehensive description of all available experimental data.
A major step forward was the development of the effective field theory of Non-Relativistic QCD (NRQCD)~\cite{Caswell:1985ui,Bodwin:1994jh}.
Within NRQCD, physical quarkonia are described as superpositions of Fock states, organized through a double expansion in the strong coupling $\alpha_s$ and in the relative velocity $v$ of the heavy-quark pair.
Production cross sections can then be factorized into perturbatively calculable Short-Distance Coefficients (SDCs), multiplied by nonperturbative Long-Distance Matrix Elements (LDMEs).
This framework provides a systematic approach to the study of quarkonium production mechanisms.
At relatively small transverse momentum, the dominant contribution arises from the direct creation of a $(Q\bar Q)$ pair in the hard scattering.
In contrast, at sufficiently large transverse momentum the fragmentation of a single high-energy parton into quarkonia becomes an increasingly important production channel~\cite{Cacciari:1994dr}.

In this work, we investigate collinear fragmentation into $S$-wave quarkonia in the color-singlet channel by employing the {\tt NRFF1.0}~\cite{Celiberto:2025euy} fragmentation functions (FFs).
These FFs are constructed within the Heavy-Flavor Non-Relativistic Evolution framework ({\HFNRevo})~\cite{Celiberto:2025euy,Celiberto:2024mex_article,Celiberto:2024bxu,Celiberto:2024rxa,Celiberto:2025xvy}, where Next-to-Leading-Order (NLO) NRQCD inputs defined at the initial scale are evolved through DGLAP dynamics and combined with a replica-based Monte Carlo treatment of missing higher-order uncertainties~\cite{Forte:2002fg}.
Within this program, FFs for conventional quarkonia provide the starting point for a broader phenomenological agenda that includes studies of heavy-flavor fragmentation in nuclear environments as well as extensions to quarkoniumlike exotic states, offering new opportunities to investigate intrinsic charm and other previously unexplored aspects of QCD dynamics.

%%%----------------------------------------
\section{Quarkonium fragmentation within {\HFNRevo}}
\label{sec:HFNrevo}
%%%----------------------------------------

\begin{figure*}[!b]
\centering

   \includegraphics[scale=0.38,clip]{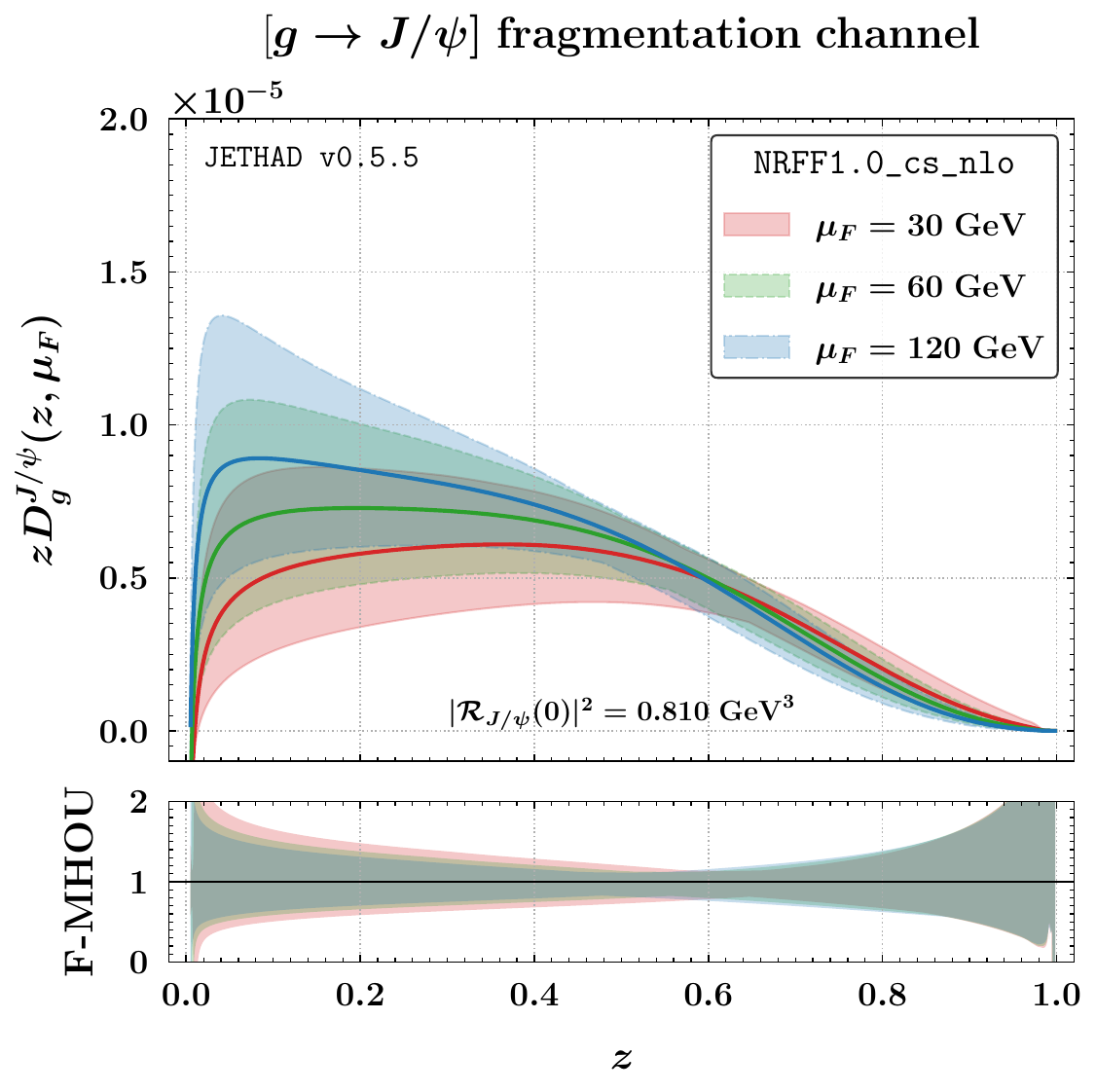}
   \hspace{0.15cm}
   \includegraphics[scale=0.38,clip]{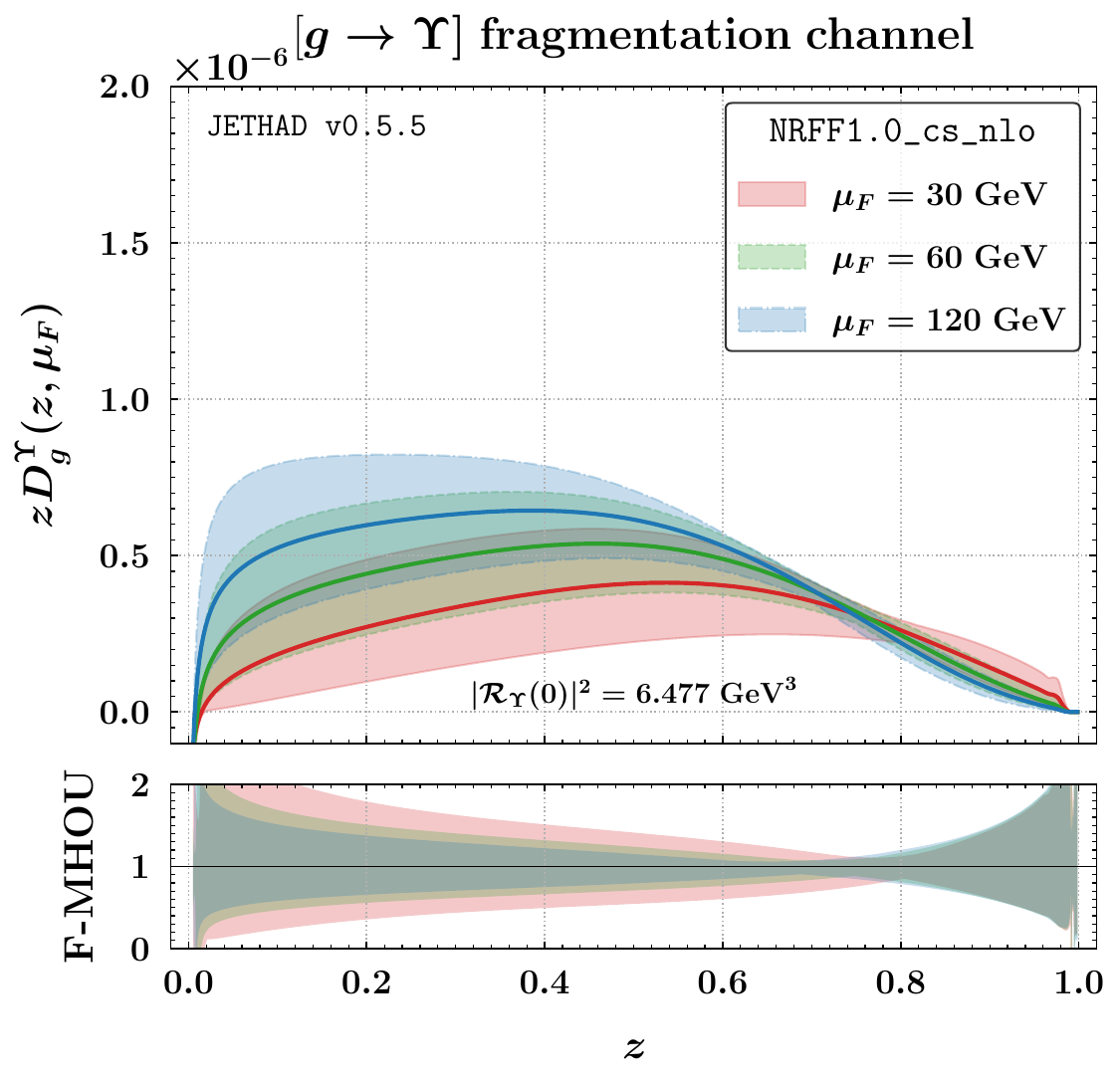}

\caption{NLO gluon to color-singlet $\Jps$ ($\Yps$) FFs. 
Plots adapted from Ref.~\protect\cite{Celiberto:2025euy}.
}

\label{fig:FFs_bottom}
\end{figure*}

\begin{figure*}[!b]
\centering

   \includegraphics[scale=0.38,clip]{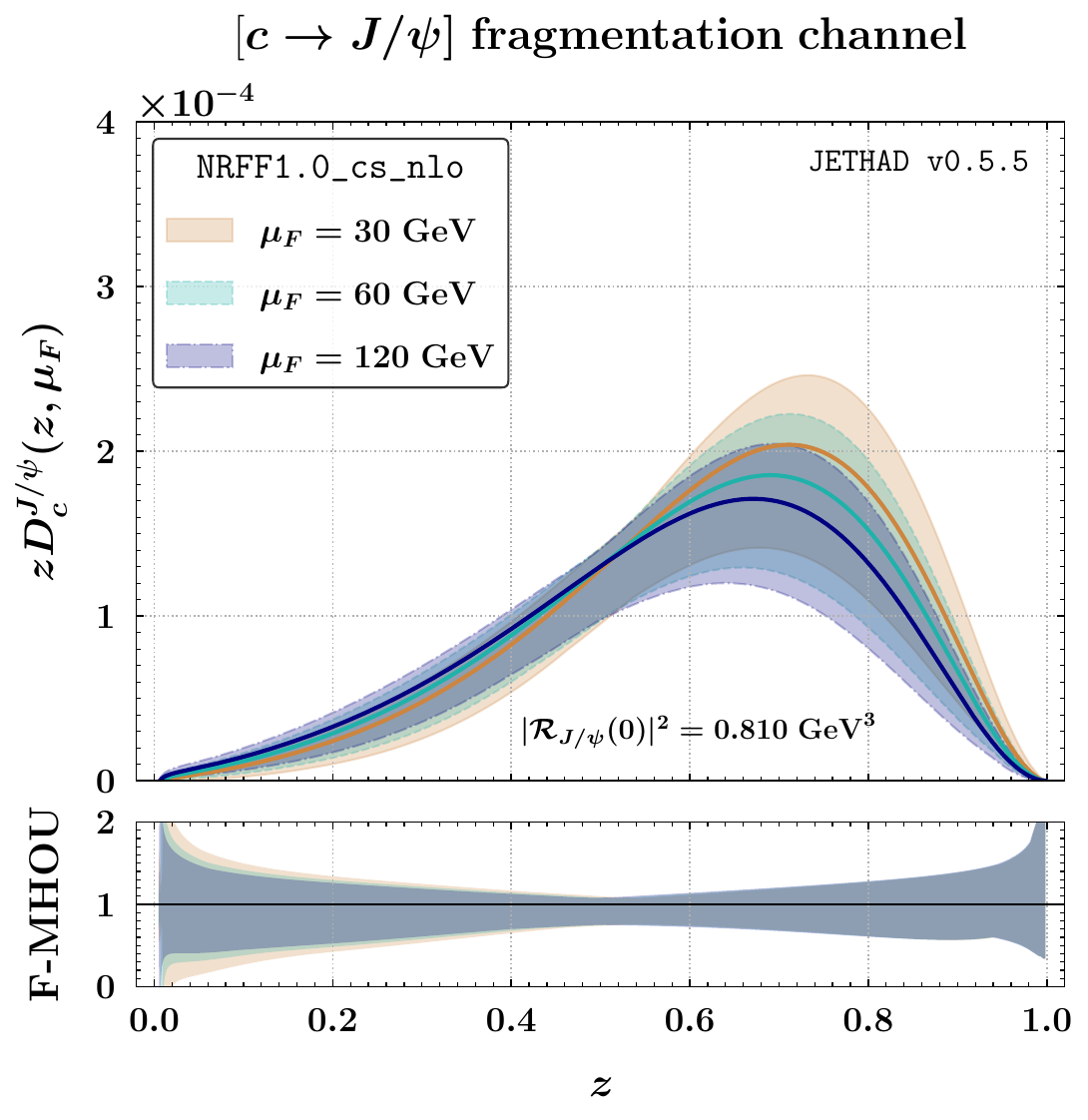}
   \hspace{0.25cm}
   \includegraphics[scale=0.38,clip]{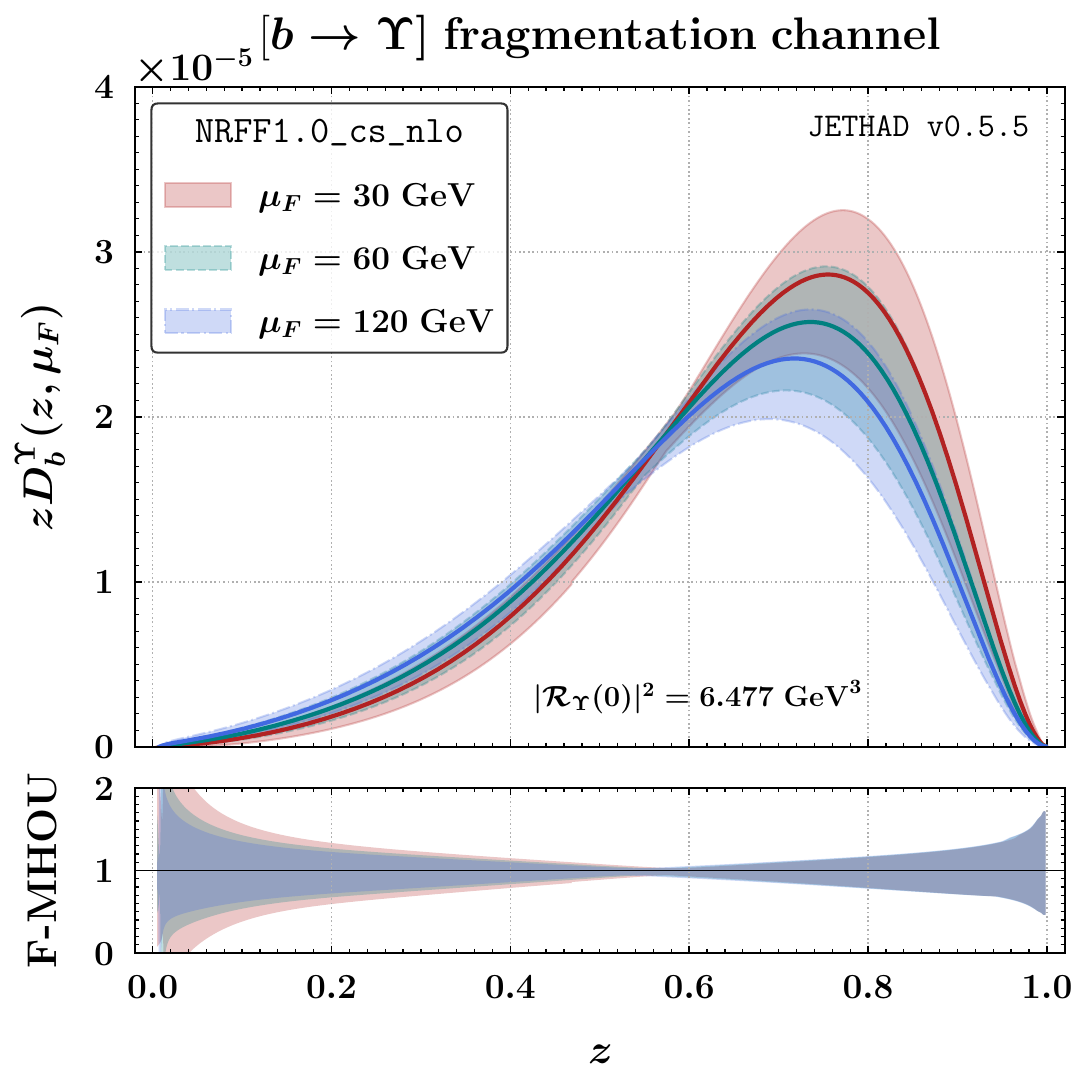}

\caption{NLO quark (bottom) to color-singlet $\Jps$ ($\Yps$) FFs. 
Plots adapted from~\protect\cite{Celiberto:2025euy}.
}

\label{fig:FFs_Q}
\end{figure*}

Because the masses of the heavy constituent quarks are significantly larger than the QCD confinement scale, $\LQCD$, the boundary conditions for heavy-flavor FFs are expected to retain a largely perturbative character.
This feature naturally supports the application of collinear factorization techniques in a fully consistent way.
To exploit this property, we have developed a new methodology, called {\HFNRevo}~\cite{Celiberto:2025euy,Celiberto:2024mex_article,Celiberto:2024bxu,Celiberto:2024rxa,Celiberto:2025xvy}, which provides a systematic framework for defining, evolving, and quantifying theoretical uncertainties in FFs for heavy quarkonia.
The {\HFNRevo} strategy is built around three conceptual pillars: physical interpretation, scale evolution, and uncertainty assessment.
From the viewpoint of interpretation, quarkonium production at relatively small transverse momentum $|\vec{q}_T|$ can be interpreted as a two-parton fragmentation mechanism within a Fixed-Flavor Number Scheme (FFNS).
This picture naturally facilitates a subsequent transition toward a Variable-Flavor Number Scheme (VFNS) formulation.
Evidence supporting this interpretation is provided by studies of transverse-momentum-dependent (TMD) observables, where distinct singular structures arise in the large-$|\vec{q}_T|$ behavior of shape functions~\cite{Echevarria:2019ynx} and in collinear FFs at intermediate transverse momentum~\cite{Boer:2023zit}.
Within {\HFNRevo}, the scale evolution of quarkonium FFs is implemented through two complementary stages.
The first step, denoted {\tt EDevo}, corresponds to a symbolic procedure in which an expanded and decoupled DGLAP evolution is constructed to properly account for heavy-flavor thresholds.
This stage is performed using the symbolic capabilities of the {\tt JETHAD} framework~\cite{Celiberto:2020wpk,Celiberto:2022rfj,Celiberto:2023fzz,Celiberto:2024mrq,Celiberto:2024swu,Celiberto:2025_P5Q_review,Celiberto:2025csa,Celiberto:2026zed} (for recent applications on high-energy precision QCD phenomenology and the proton structure at low-$x$, see Refs.~\cite{Celiberto:2015yba,Celiberto:2015mpa,Celiberto:2016ygs,Celiberto:2016vva,Caporale:2018qnm,Celiberto:2022gji,Celiberto:2016hae,Celiberto:2016zgb,Celiberto:2017ptm,Celiberto:2017uae,Celiberto:2017ydk,Bolognino:2018oth,Bolognino:2019cac,Bolognino:2019yqj,Celiberto:2020rxb,Celiberto:2022kxx,Celiberto:2020tmb,Celiberto:2023rtu,Celiberto:2023uuk_article,Celiberto:2023eba,Celiberto:2023nym,Celiberto:2023dkr_article,Celiberto:2023rqp,Celiberto:2024mdt,Celiberto:2024bfu,Celiberto:2025edg,Celiberto:2021fjf,Celiberto:2021tky,Celiberto:2021txb,Celiberto:2021xpm,Bolognino:2021mrc,Bolognino:2021hxx,Celiberto:2017nyx,Bolognino:2019ouc,Bolognino:2019yls,Bolognino:2019ccd,Celiberto:2021dzy,Celiberto:2021fdp,Bolognino:2022wgl,Celiberto:2022dyf,Celiberto:2022grc,Bolognino:2022paj,Celiberto:2022qbh,Celiberto:2022keu,Celiberto:2022zdg,Celiberto:2022kza,Celiberto:2024omj,Celiberto:2025euy,Celiberto:2017ius,Mohammed:2022gbk,Gatto:2025kfl} and~\cite{Bolognino:2018rhb,Bolognino:2018mlw,Bolognino:2019bko,Bolognino:2019pba,Celiberto:2019slj,Bolognino:2021niq,Bolognino:2021gjm,Bolognino:2022uty,Celiberto:2022fam,Bolognino:2022ndh,Celiberto:2018muu,Bolognino:2021bjd}, respectively).
The second stage, labeled {\tt AOevo}, performs the full numerical DGLAP evolution to all orders in the resummed logarithmic structure.
A further essential ingredient of {\HFNRevo} is the systematic treatment of Missing Higher-Order Uncertainties (MHOUs) associated with the choice of evolution thresholds.
In practice, these effects are explored through correlated variations of renormalization and factorization scales entering FF initial inputs.
The two scales are varied simultaneously by a factor of two around their central values.
This strategy follows the spirit of recent approaches developed for PDF theory uncertainties, including theory-covariance-matrix techniques~\cite{NNPDF:2024dpb} and the {\tt MCscales} procedure~\cite{Kassabov:2022orn}.
To illustrate the phenomenological behavior of quarkonium FFs within this framework, we analyze four representative fragmentation channels calculated at NLO accuracy~\cite{Braaten:1993rw,Braaten:1993mp,Artoisenet:2014lpa,Zhang:2018mlo,Zheng:2021ylc,Zheng:2021mqr}.
Results for the pseudoscalar states $\ecs(^1S_0^{(1)})$ and $\ebs(^1S_0^{(1)})$ were presented in Ref.~\cite{Celiberto:2025euy}, while in this work we report preliminary results for the vector quarkonia $\Jps(^3S_1^{(1)})$ and $\Yps(^3S_1^{(1)})$.
Figure~\ref{fig:FFs_bottom} (left and right panels) shows the {\tt NRFF1.0} gluon-induced FFs for $(g \to \Jps)$ and $(g \to \Jps)$, evolved with $\mu_F$ spanning the same range considered in previous analyses.
Similarly, Fig.~\ref{fig:FFs_Q} displays the charm-induced FFs for $(c \to \Jps)$ and $(b \to \ebs)$, with $\mu_F$ varied between 30 and 120~GeV.

%%%----------------------------------------
\section{Quarkonium-in-jet fragmentation}
\label{sec:onium_in_jet}
%%%----------------------------------------

Jet substructure measurements have recently become an essential tool for probing the underlying dynamics of the strong interaction.
Beyond improving our understanding of QCD, these observables also provide promising opportunities to search for possible signals of physics beyond the Standard Model.
A systematic analysis of the internal composition of jets, particularly when heavy-flavored hadrons are identified among their constituents, offers valuable information on both perturbative and nonperturbative mechanisms governing QCD dynamics~\cite{Procura:2009vm,Bauer:2013bza,Chien:2015ctp,Maltoni:2016ays,Kang:2017glf,Metodiev:2018ftz,Marzani:2019hun,Kasieczka:2020nyd,Nachman:2022emq,Dhani:2024gtx}.
A relevant class of observables concerns the identification of a specific hadron within the reconstructed jet.
Within the framework of collinear factorization, this situation is described through the formalism of Semi-Inclusive Fragmenting Jet Functions (SIFJFs).
At leading power, the SIFJF describing the fragmentation of a parton $i$ into a quarkonium state ${\cal H}_Q$ observed inside the jet can be written as~\cite{Kang:2017yde}
\begin{equation}
 \label{eq:SIFJF}
 {\cal F}_i^{\cal H}(z, z_{\cal H}, \mu_F, R_J) \, = \,
 \sum_{j=q,\bar{q},g} \int_{z_{\cal H}}^1 \frac{\drv \zeta}{\zeta} \,
 {\cal S}(z, z_{\cal H}/\zeta, \mu_F, R_J) \,
 D_j^{\cal H}(\zeta, \mu_F) \;,
\end{equation}
where $D_j^{\cal H}(\zeta, \mu_F)$ denotes the conventional $(j \to {\cal H}_Q)$ collinear fragmentation channel.
Furthermore, ${\cal S}(z, z_{\cal H}/\zeta, \mu_F, R_J)$ are the perturbatively calculable fragmenting jet coefficients~\cite{Baumgart:2014upa}, which are currently known at NLO accuracy for both anti-$\kappa_T$ and cone jet algorithms~\cite{Kang:2016ehg}.
In Eq.~\ref{eq:SIFJF}, the variable $z$ corresponds to the ratio between the light-cone momentum carried by the reconstructed jet and that of the initiating parton $i$.
Analogously, $z_{\cal H}$ is defined as the fraction of the jet light-cone momentum taken by the identified hadron.
The parameter $R_J$ denotes the jet radius.

%%%----------------------------------------
\section{Bridging to heavy ions}
\label{sec:heavy_ions}
%%%----------------------------------------

Ultra-relativistic heavy-ion collisions provide a unique environment for the investigation of QCD phenomena under extreme conditions, where a deconfined Quark-Gluon Plasma (QGP) is formed and modifies the propagation and hadronization of heavy quarks. 
In this setting, quarkonium production is significantly altered with respect to the proton-proton baseline, leading to a characteristic suppression pattern observed for both charmonium and bottomonium states. 
This suppression can be traced back to two primary mechanisms~\cite{Vogt:1999cu,Satz:2000bn}. 
The first is \emph{color screening}, whereby the confining potential between the heavy quark and antiquark is weakened by the presence of thermal color charges in the medium~\cite{Shuryak:1980tp,Heinz:2000bk,Braun-Munzinger:2015hba}. 
This effect induces a sequential melting of quarkonium states, with more weakly bound states dissociating at lower temperatures. 
Consequently, tightly bound ground states such as $J/\psi$ and $\Upsilon$ exhibit greater resilience than excited states like $\psi(2S)$ and $\Upsilon(2S)$, as well as their pseudoscalar counterparts $\eta_{c}$ and $\eta_{b}$~\cite{Matsui:1986dk,Grandchamp:2003uw,GayDucati:2003xa}.
A second key mechanism is gluon-induced \emph{dissociation}, in which thermal gluons interact inelastically with quarkonium states and promote their breakup. 
At sufficiently high collision energies, this suppression channel is partially compensated by quarkonium \emph{regeneration}, arising from the recombination of independently produced heavy quarks within the medium. 
The observed yields therefore result from a competition between dissociation and regeneration, with their balance controlled by the temperature, lifetime, and heavy-quark density of the QGP~\cite{Andronic:2008gu,Grandchamp:2003uw}.
A quantitative description of these phenomena typically relies on transport approaches and hydrodynamic simulations, supplemented by lattice-QCD inputs for in-medium binding energies and dissociation rates. 
Despite these advances, sizable uncertainties persist, particularly in the high-transverse-momentum regime and for pseudoscalar states such as $\eta_c$ and $\eta_b$.

Within this context, the {\HFNRevo} framework provides a well-defined vacuum baseline for the construction of medium-modified FFs for quarkonia. 
Thanks to its scale-differential evolution and explicit treatment of heavy-flavor thresholds, {\HFNRevo} can be systematically extended to incorporate parton energy loss, quenching effects, and medium-induced distortions of fragmentation patterns in nuclear environments. 
It also offers a consistent starting point for embedding in-medium hadronization mechanisms, including statistical recombination and color de-excitation processes.
In addition, {\HFNRevo} enables the definition of observables sensitive to modifications of fragmentation dynamics in the QGP. 
In particular, the notion of fragmentation-function apparent-shape distortion (FF-ASD) provides a convenient parameterization of deviations from vacuum FFs induced by thermal broadening, dissociative effects, or delayed hadron formation. 
Such observables can be accessed through high-transverse-momentum quarkonium-tagged jet measurements at the HL-LHC and future collider facilities, opening new avenues for tomographic studies of the QGP.

%%%----------------------------------------
\section{Quarkoniumlike States and Intrinsic Charm}
\label{sec:IC}
%%%----------------------------------------

Early leading-order calculations for gluon- and heavy-quark-initiated fragmentation into $S$-wave vector quarkonia~\cite{Braaten:1993rw,Braaten:1993mp} were subsequently improved through next-to-leading-order corrections~\cite{Zheng:2019gnb,Zheng:2021sdo}. 
These developments enabled the construction of the first VFNS-evolved FF sets~\cite{Celiberto:2022dyf,Celiberto:2023fzz}, together with their charmed $B$-meson extensions~\cite{Celiberto:2022keu,Celiberto:2024omj}. 
The resulting predictions showed good agreement with LHCb measurements~\cite{LHCb:2014iah,LHCb:2016qpe,Celiberto:2024omj}, confirming the expected suppression of $B_c$ production relative to $B$ mesons at the level of $\sim 0.1\%$~\cite{Celiberto:2024omj}, and providing strong evidence for the validity of the VFNS approach in the high-transverse-momentum regime.

The NRQCD-based description of fragmentation has progressively been generalized to encompass exotic multiquark states. 
Experimental observations of double $J/\psi$ final states~\cite{LHCb:2020bwg,ATLAS:2023bft,CMS:2023owd} have been interpreted as signals of compact tetraquark configurations~\cite{Zhang:2020hoh,Zhu:2020xni}. 
In this picture, the production mechanism involves the short-distance creation of two heavy-quark pairs, followed by their nonperturbative binding into a multiquark state. 
The first NRQCD-based inputs for gluon fragmentation into color-singlet $S$-wave $T_{4c}$ states were derived in~\cite{Feng:2020riv}. 
Building on these results, the {\tt TQHL1.0} FFs~\cite{Celiberto:2023rzw,Celiberto:2024mrq} provided the first VFNS-consistent description of heavy-light tetraquark fragmentation. 
Subsequent developments, including the {\tt TQ4Q1.x}~\cite{Celiberto:2024mab,Celiberto:2025dfe,Celiberto:2025ziy} and {\tt TQHL1.1}~\cite{Celiberto:2024beg} sets, introduced NRQCD-driven modeling for both gluon~\cite{Feng:2020riv} and heavy-quark~\cite{Bai:2024ezn} initiated channels, improved the treatment of doubly heavy configurations, and extended the framework to bottomoniumlike states.

A dedicated analysis of axial-vector ($1^{+-}$) tetraquarks was presented in~\cite{Celiberto:2025dfe}, where uncertainties associated with LDMEs were consistently propagated to the FFs. 
The {\tt TQ4Q1.1} framework was later employed to explore phenomenological signatures of charmed tetraquarks in Higgs and electroweak decay channels~\cite{Ma:2025ryo}. 
In parallel, FF constructions were extended to other classes of exotic hadrons, including fully charmed pentaquarks ({\tt PQ5Q1.0}) and triply heavy $\Omega$ baryons ({\tt OMG3Q1.0})~\cite{Celiberto:2025ipt,Celiberto:2025ogy}, thereby establishing a unified strategy for the study of heavy exotic production within QCD.

A further step in this direction has recently been achieved in Ref.~\cite{Celiberto:2025vra}, where the production of fully heavy tetraquarks has been investigated in the forward-rapidity region. 
While previous analyses primarily focused on central rapidities, this study explores asymmetric kinematic configurations characterized by $x_1 \gg x_2$, which enhance sensitivity to both small-$x$ non-linear dynamics and the large-$x$ structure of the projectile. 
The corresponding dilute-dense scattering regime is described within a hybrid-factorization framework, combining collinear inputs with a gluon-saturation-based treatment of the target. 
Both gluon-initiated and charm-initiated production channels are included, together with different models for the intrinsic component of the proton wave function.

The resulting phenomenology shows that tensor tetraquark states are predominantly produced through gluon fragmentation, yielding the largest cross sections across a wide kinematic range, from the 13~TeV LHC to prospective 100~TeV FCC energies~\cite{FCC:2025lpp,FCC:2025uan,FCC:2025jtd}. 
In contrast, axial-vector states are dominantly generated via charm-initiated fragmentation, making them especially sensitive to the large-$x$ charm content of the proton. 
In this respect, the production of $T_{4c}(1^{+-})$ states emerges as a particularly clean probe of intrinsic charm, especially in forward kinematics where charm-initiated contributions are significantly enhanced.

%%%----------------------------------------
\section{Conclusions and Outlook}
\label{sec:conclusions}
%%%----------------------------------------

We have presented the {\HFNRevo} framework as a novel and systematically improvable approach to construct quarkonium FFs in the collinear regime.
Within this setup, we have introduced the first release of the {\tt NRFF1.0} set~\cite{Celiberto:2025euy}, which provides color-singlet initial-scale inputs for all partonic channels, derived from NLO NRQCD calculations.
The subsequent scale evolution is performed through DGLAP equations with a consistent implementation of heavy-flavor thresholds, while theoretical uncertainties are quantified via a Monte Carlo replica-like procedure designed to capture missing higher-order contributions.
The {\tt NRFF1.0} set~\cite{Celiberto:2025euy} represents a significant advancement over previous determinations such as {\tt ZCW19}$^+$ and {\tt ZCFW22}, which have been employed in phenomenological analyses of vector quarkonia~\cite{Celiberto:2022dyf,Celiberto:2023fzz} and charmed $B$-meson production~\cite{Celiberto:2022keu,Celiberto:2024omj}.
Owing to its improved theoretical consistency, this framework is particularly well suited for precision studies at current and future experimental facilities, including the HL-LHC~\cite{Chapon:2020heu,Amoroso:2022eow,LHCspin:2025lvj}, the EIC~\cite{AbdulKhalek:2021gbh,Khalek:2022bzd,Abir:2023fpo} and next-generation colliders~\cite{AlexanderAryshev:2022pkx,LinearColliderVision:2025hlt,LinearCollider:2025lya,Gessner:2025acq,Black:2022cth,Accettura:2023ked,InternationalMuonCollider:2024jyv,MuCoL:2024oxj,MuCoL:2025quu,InternationalMuonCollider:2025sys}, as well as for benchmarking data-driven and machine-learning-based FF extractions~\cite{Allaire:2023fgp,Hammou:2023heg,Costantini:2024xae}.
Future developments of the {\HFNRevo} program include the incorporation of color-octet contributions for vector quarkonia~\cite{Cho:1995vh,Cacciari:1996dg}, the implementation of a general-mass VFNS~\cite{Cacciari:1998it,Forte:2010ta,Guzzi:2011ew}, and systematic applications to rare~\cite{Celiberto:2025ogy} and exotic heavy hadrons~\cite{Celiberto:2023rzw,Celiberto:2024mrq,Celiberto:2024mab,Celiberto:2024beg,Celiberto:2025dfe,Celiberto:2025ziy,Celiberto:2025vra,Celiberto:2025ipt}.
In particular, the extension of this framework to fully heavy tetraquarks has opened a new phenomenological direction, where specific channels, such as axial-vector configurations, can act as sensitive probes of the intrinsic charm content of the proton in forward kinematics.
A longer-term objective is the extension of {\HFNRevo} to quarkonium-in-jet fragmentation, enabling detailed studies of jet substructure through quarkonium-tagged observables and resummation-sensitive angular correlations.
At the same time, {\HFNRevo} provides a robust baseline for investigating medium-induced effects on heavy-flavor fragmentation in nuclear collisions.
Its threshold-resolved structure and explicit treatment of partonic hierarchies make it particularly suitable for studying jet quenching, parton energy loss, and the emergence of medium-modified fragmentation functions in the quark-gluon plasma.
It also offers a consistent framework to incorporate in-medium hadronization mechanisms, including regeneration processes and FF-ASDs, thereby defining new observables for the tomographic exploration of the QGP at the HL-LHC and future collider facilities.

%%%----------------------------------------
\section*{Acknowledgments}
\label{sec:acknowledgments}
%%%----------------------------------------

We are supported by the Atracci\'on de Talento Grant n. 2022-T1/TIC-24176 (Madrid, Spain).

%-----------------------------------------
\vspace{-0.05cm}
\begingroup
\setstretch{0.6}
\bibliographystyle{bibstyle}
\bibliography{biblography}
\endgroup
%-----------------------------------------

\end{document}